\documentclass[12pt]{article}
\pdfoutput=1
\usepackage{amsmath,amsfonts,amssymb,graphicx,float}
\usepackage[top=3cm, bottom=3cm, left=2.5cm, right=2.5cm]{geometry}

\numberwithin{equation}{section}

\usepackage{perpage}
\MakePerPage[2]{footnote}

\usepackage[hyperfootnotes=false,bookmarks=true]{hyperref}
\makeatletter
\renewcommand{\eqref}[1]{\textup{\tagform@{\ref*{#1}}}}
\makeatother

\DeclareMathOperator\arctanh{arctanh}
\DeclareMathOperator\sech{sech}

\begin{document}

\pagestyle{empty}

\hfill UMN-TH-3407/14

\begin{center}
\vspace{1cm}
{\Large\bf A Soft-Wall Dilaton}\\

\vspace{1cm}
{\sc Peter Cox$^{a,}$\footnote{Email: pcox@physics.unimelb.edu.au} {\small and}
Tony Gherghetta$^{b,}$\footnote{Email: tgher@umn.edu}}
\vspace{0.5cm}

{\it \small {$^a$ARC Centre of Excellence for Particle Physics at the Terascale,\\
School of Physics, University of Melbourne, Victoria 3010, Australia}}

{\it \small {$^b$School of Physics \& Astronomy, University of Minnesota, \\Minneapolis, MN 55455, USA}}\\
\end{center}

\vspace{1cm}
\begin{abstract}
\baselineskip=15pt
\noindent
We study the properties of the dilaton in a soft-wall background using two solutions of the Einstein equations. These solutions contain an asymptotically AdS metric with a nontrivial scalar profile that causes both the spontaneous breaking of conformal invariance and the generation of a mass gap in the particle spectrum. We first present an analytic solution, using the superpotential method, that describes a CFT spontaneously broken by a finite dimensional operator in which a light dilaton mode appears in the spectrum. This represents a tuning in the vanishing of the quartic coupling in the effective potential that could be naturally realised from an underlying supersymmetry. Instead, by considering a generalised analytic scalar bulk potential that quickly transitions at the condensate scale from a walking coupling in the UV to an order-one $\beta$-function in the IR, we obtain a naturally light dilaton. This provides a simple example for obtaining a naturally light dilaton from nearly-marginal CFT 
deformations in the more realistic case of a soft-wall background.

\end{abstract}

\newpage
\pagestyle{plain}

%%%%%%%%%%%%%%%%%%%%%%%%%%%%%%%%%%%%%%%%%%%%%%%%%%%%%%%%%%%%%%%%%%%%%%%%%%%%%%%%%%%%%%%%%%%%%%%%%%%
\section{Introduction}

Conformal invariance is a nontrivial spacetime symmetry which provides an alternative way to address the question of naturalness in field theory. When the scale (or dilation) symmetry is spontaneously broken, the Nambu-Goldstone boson or dilaton is a massless scalar particle protected by a shift symmetry.  However unlike the spontaneous breaking of a global symmetry, the non-linearly realised scale symmetry allows for the presence of a quartic self-interaction of the dilaton. This is incompatible with the symmetry breaking unless the quartic coupling vanishes, either by a fine-tuning or due to an additional symmetry, such as supersymmetry. This implies that generically in conformal field theories (CFTs), conformal invariance is not spontaneously broken and  there is no massless dilaton.

A way to avoid requiring special conditions is to introduce a small amount of explicit breaking of conformal symmetry. In this case the dilaton obtains a mass but can remain naturally light. The conditions under which this is possible were considered by Contino, Pomarol and Rattazzi~\cite{CPR}. By assuming that conformal invariance is explicitly broken by a Lagrangian deformation $\lambda {\cal O}$,  a naturally light dilaton occurs  for a near-marginal operator $\cal O$, so that the running coupling $\lambda(\mu)$ remains close to marginality throughout the renormalisation-group (RG) evolution. Similar effects were also studied in the four-dimensional (4D) effective theory in Ref~\cite{Chacko:2012sy}. Explicit holographic realisations were subsequently given in \cite{Bellazzini:2013fga, Coradeschi:2013gda} where the near-marginal deformation of the CFT corresponds to introducing a nearly massless bulk scalar field. The approximately constant bulk scalar potential preserves an approximate shift symmetry 
which leads to a renormalisation-group flow with a small $\beta$-function. However to obtain a dilaton in the low-energy spectrum, these explicit realisations also assumed the presence of an IR brane (or hard wall) which corresponds to spontaneously breaking the conformal symmetry by another operator which has an arbitrarily large dimension. Thus by introducing two scalar operators there is a simple, although idealised way to obtain a light dilaton.

A more realistic framework to study the properties of the dilaton is to consider a soft-wall background. This corresponds to introducing a single bulk scalar field with a nontrivial bulk scalar potential. The solutions of the coupled Einstein-scalar equations of motion can then lead to a scalar profile that grows in the IR, causing a back-reaction on the metric that deviates from AdS. Thus the ``soft wall" produced by the single scalar field causes the spontaneous breaking of conformal symmetry with a mass gap in the spectrum. Equivalently, in the holographic description, there is a single operator $\cal O$ that is responsible for explicitly breaking the conformal symmetry and generating a condensate $\langle \cal O \rangle$. The fluctuations about the condensate $\langle \cal O \rangle$ are then identified with the dilaton. 

Nearly-marginal deformations in a soft-wall background can still lead to a naturally light dilaton and were first considered in Ref.~\cite{Megias:2014iwa}. As long as the holographic $\beta$-function remains small at the condensation scale and then transitions sufficiently quickly to an order-one constant in the IR, an approximate shift symmetry can be retained. This translates into requiring a bulk scalar potential that must transform from an approximately constant potential near the UV brane to an exponential potential in the IR. If the transition region is sufficiently small then an approximate shift symmetry can be preserved with a corresponding light dilaton. In Ref.~\cite{Megias:2014iwa} an approximate piecewise solution was constructed that exhibited these features in a soft-wall background and an approximate mass formula for the dilaton was derived.

In this paper we present soft-wall solutions that provide a simple description of the dilaton from the UV to the IR scale. Using the superpotential method we first derive an analytic  solution of the coupled Einstein-scalar equations of motion with an asymptotically AdS metric and nontrivial scalar profile that grows in the IR. This solution parametrises the explicit and spontaneous breaking of conformal symmetry by operators with dimensions in the range $1<[{\cal O}]<4$. A light dilaton is obtained when there is spontaneous breaking of the conformal symmetry and a hierarchy between the UV and IR scales. This is similar to the result obtained in Ref.~\cite{Gherghetta:2011rr} where a general analysis of scalar fluctuations in a soft-wall background was given. In addition, we show that a light dilaton is obtained in the case of explicit breaking by an operator of dimension $[{\cal O}] \simeq 2$. However, in both cases this corresponds to a tuning in the quartic coupling of the dilaton effective potential. This 
is a consequence of using an analytic superpotential and presumably the tuning could be understood from an underlying supersymmetry.

The scalar potential derived from the superpotential is then generalised to allow for non-analytic terms in the $\beta$-function. This allows for a deformed CFT with a nonzero condensate $\langle \cal O\rangle$. We find that a naturally light dilaton only occurs for nearly-marginal operators. In this case the $\beta$-function is approximately constant in the UV and then transitions rapidly to an approximately order-one constant in the IR with the corresponding scalar potential respecting an approximate shift symmetry. This agrees with the results derived in Ref.~\cite{Megias:2014iwa}. Furthermore with our solution we quantify how fast the rise to confinement must be in order to obtain a naturally light dilaton. We also find that in our case the $\beta$-function has a steep slope in the transition region precisely when the operator is close to marginality.

The outline for the paper is as follows: In Section~\ref{sec:solution} we review the soft-wall background and the holographic interpretation. An analytic superpotential solution is presented in Section~\ref{sec:analytic_solution} where we give the conditions for a discrete spectrum and calculate the dilaton mass. A light dilaton occurs when conformal invariance is spontaneously broken and there is a hierarchy between the UV and IR scales. Alternatively in the case when there is an explicit UV breaking a light dilaton occurs when the operator dimension is near two. Both cases correspond to tuned scenarios. In Section~\ref{sec:general_case} we consider a generalised analytic bulk scalar potential. In this case we show that a naturally light dilaton does occur but only for nearly-marginal operators. Our conclusion is given in Section~\ref{sec:conclusion}.

%%%%%%%%%%%%%%%%%%%%%%%%%%%%%%%%%%%%%%%%%%%%%%%%%%%%%%%%%%%%%%%%%%%%%%%%%%%%%%%%%%%%%%%%%%%%%%%%%%%
\section{The Soft-Wall Solution} \label{sec:solution}

We begin by considering a five-dimensional (5D) spacetime labelled by the coordinates $(x^\mu, y)$ with $\mu=0,1,2,3$. The general 5D action is given by
\begin{equation} \label{eq:action}
 S=2\int^{y_c}_{y_0}d^5x\,\sqrt{-g}\left(\frac{M^3}{2}R-\frac{1}{2}(\partial\phi)^2-V(\phi)\right)-\int_{UV}d^4x\,\sqrt{-\gamma}\left(2M^3[K]+U(\phi)\right),
\end{equation}
where $M$ is the 5D Planck scale and $V(\phi)$ is the bulk scalar potential.  A single 4D brane located at $y_0$ provides a UV cut-off, while $y_c$ represents a curvature singularity. The UV brane has an induced metric $\gamma$, a brane potential $U(\phi)$ and $[K]$ represents the jump in the extrinsic curvature. The requirement of 4D Poincar\'{e} invariance leads to the usual metric ansatz
\begin{equation} \label{eq:bkg_metric}
 ds^2=g_{MN} dx^M dx^N=A^2(y)\eta_{\mu\nu}dx^{\mu}dx^\nu+dy^2,
\end{equation}
where the indices $M,N=(\mu,5)$, $A(y)$ is the warp factor and $\eta_{\mu\nu} = {\rm diag}(-1,+1,+1,+1)$ is the 4D Minkowski metric.  The scalar profile $\phi=\phi(y)$ will in general be a nontrivial function of the 5th coordinate, $y$. The functions $A(y)$ and $\phi(y)$ therefore characterise the soft-wall background.

As usual, it is convenient to express the Einstein equations as a simple first-order system in terms of a superpotential, $W$, defined by~\cite{DeWolfe:1999cp}
\begin{equation} \label{eq:superpotential_def}
 V(\phi)=\frac{1}{2}\,W'^2(\phi)-\frac{2}{3M^3}\,W^2(\phi).
\end{equation}
The bulk equations of motion for the metric and scalar profile are then simply given by
\begin{equation} \label{eq:bulk_bkg_eom}
 \phi'(y)=W'(\phi), \qquad \frac{A'(y)}{A(y)}=-\frac{1}{3M^3}\,W(\phi),
\end{equation}
with the additional conditions evaluated at the UV boundary
\begin{equation} \label{eq:bdry_bkg_eom}
 W(\phi)=\frac{1}{2}\,U(\phi), \qquad W'(\phi)=\frac{1}{2}\,U'(\phi).
\end{equation}

%%%%%%%%%%%%%%%%%%%%%%%%%%%%%%%%%%%%%%%%%%%%%%%%%%%%%%%%%%%%%%%%%%%%%%%%%%%%%%%%%%%%%%%%%%%%%%%%%%%
\subsection{Holographic $\beta$-function} \label{sec:beta_function}

In order to investigate the behaviour of a given soft-wall solution and its interpretation in terms of a 4D CFT we shall make use of the holographic $\beta$-function \cite{Behrndt:1999ay,deBoer:1999xf,Anselmi:2000fu}. As usual, the AdS/CFT correspondence allows us to relate the warp factor to the energy scale, $\mu$ in the 4D theory according to 
\begin{equation}
 \mu=kA(y)\simeq k\,e^{-ky}. 
\end{equation}
It is well known in the case of RS models with a Goldberger-Wise field that the UV value of the bulk scalar can be identified with the UV value of a coupling responsible for deforming the CFT~\cite{Rattazzi:2000hs}. The definition of the holographic $\beta$-function, $\beta(\phi)$ then immediately follows:
\begin{equation} \label{eq:beta_def}
 \beta(\phi)\equiv A\partial_A\phi=-3M^3\frac{W'(\phi)}{W(\phi)},
\end{equation}
where we have used the bulk equations \eqref{eq:bulk_bkg_eom}. The above equation can be solved to obtain the superpotential in terms of $\beta$ and so a given model can be equivalently defined in terms of either the $\beta$-function or superpotential. The most common situation is to consider $\beta$-functions possessing a UV fixed point, corresponding to a bulk potential with an AdS extremum. However, it is still possible to obtain a wide variety of RG flows and spectra depending on the behaviour of the $\beta$-function in the IR. This has been discussed in detail in \cite{Cabrer:2009we} in terms of the superpotential and also recently in \cite{Megias:2014iwa}. 

We shall be interested in cases which result in a confining geometry with a discrete spectrum; this translates into a limit on the asymptotic value of the $\beta$-function as $\phi\rightarrow\infty$,
\begin{equation} \label{eq:beta_limit}
 \sqrt{3M^3}<-\beta_\infty<2\sqrt{3M^3}.
\end{equation}
The lower bound arises from the requirement that the theory is confining in the IR\footnote{The case -$\beta_\infty=\sqrt{3}$ is marginal and depends on the detailed behaviour of the $\beta$-function as $\phi\rightarrow\infty$ \cite{Cabrer:2009we}.}. The upper bound is obtained after imposing boundary conditions in the IR. The simplest condition is to require that the solution satisfies the Einstein equations at the dynamically generated boundary \cite{Cabrer:2009we}. Alternatively AdS/CFT can be used to restrict to physical singularities with a finite temperature field theory dual~\cite{Gubser:2000nd}; in either case the same restriction on the $\beta$-function is obtained. For a given bulk potential, imposing these IR boundary conditions fixes the integration constant in the solution for the superpotential (or $\beta$-function) in Eq.~\eqref{eq:superpotential_def}.

%%%%%%%%%%%%%%%%%%%%%%%%%%%%%%%%%%%%%%%%%%%%%%%%%%%%%%%%%%%%%%%%%%%%%%%%%%%%%%%%%%%%%%%%%%%%%%%%%%%
\subsection{UV Behaviour}

In order to characterise the type of RG flow described by a given $\beta$-function it is useful to consider the behaviour of the solutions near the UV fixed point. Taking a bulk potential with an AdS extremum, which we assume to be at $\phi=0$ without loss of generality, and expanding around this point we have
\begin{equation} \label{eq:UV_bulk_potential}
 V(\phi)=-6k^2M^3+\frac{1}{2}m_\phi^2 \phi^2+O(\phi^3)\,,
\end{equation}
where $k$ is the AdS curvature scale and $m_\phi$ is the scalar bulk mass. Solving Eq.~\eqref{eq:superpotential_def} the superpotential then takes the form\footnote{In the case $\Delta_+=\Delta_-=2$, the $\Delta_-$-type superpotential is $W(\phi)\simeq 3kM^3 +k\,\phi^2\left(1+\frac{1}{\log{\phi}}\right)$ and $\phi(A)\simeq\lambda\,A^{-2}\log(A)+\langle\mathcal{O}\rangle\, A^{-2}$.}
\begin{equation} \label{eq:UV_superpotential}
 W(\phi)=3kM^3+\frac{1}{2}k\Delta_\pm\phi^2\,+\, \ldots\,,
\end{equation}
where the $\Delta_\pm$ are given in terms of the bulk mass by $\Delta_\pm=2\pm\sqrt{4+m_\phi^2/k^2}$. For a given bulk potential there are therefore two possible asymptotic forms for the superpotential, which shall be referred to as $\Delta_+$ and $\Delta_-$-type. It was shown in Ref.~\cite{Papadimitriou:2007sj} that $\Delta_-$-type solutions also admit a leading non-perturbative term, $\phi^{4/\Delta_-}$, resulting in a family of $\Delta_-$-type solutions parameterised by the integration constant in Eq.~\eqref{eq:superpotential_def}. Such a term is forbidden for $\Delta_+$-type solutions, which are either isolated or located at an infinite distance in parameter space from any $\Delta_-$-type solution.

From Eq.~\eqref{eq:bulk_bkg_eom} and using the fact that the metric is asymptotically AdS the solution for the bulk scalar in terms of the warp factor is 
\begin{equation} \label{eq:UV_scalar_vev}
 \phi(A)=\lambda\, A^{-\Delta_-}+\frac{\langle\mathcal{O}\rangle}{2\Delta_+-4}\, A^{-\Delta_+} \,+\, \ldots\,.
\end{equation}
Assuming Dirichlet UV boundary conditions for the bulk scalar, the two leading terms can then be related to the running coupling and the condensate of the deforming operator in the CFT~\cite{Klebanov:1999tb}. The $\Delta_\pm$ are related to the dimension of the operator according to $[\mathcal{O}]=\Delta_+=4-\Delta_-$, leading to the standard AdS/CFT relation between the operator dimension and bulk mass. Provided that $1\leq\Delta_-\leq2$, one can also consider Neumann or mixed boundary conditions in which case we have an alternative CFT description where the leading term is now identified with the condensate and the operator dimension is given by $\Delta_-$.

Considering more carefully the two types of superpotential solutions, it is clear that in the case of $\Delta_+$-type solutions the leading term in Eq.~\eqref{eq:UV_scalar_vev} will be absent and therefore these solutions correspond to a CFT with purely spontaneous breaking of the conformal symmetry (assuming Dirichlet boundary conditions). On the other hand, $\Delta_-$-type solutions correspond to a deformed CFT and one finds that the coefficients of the two terms are in fact related according to
\begin{equation} \label{eq:condensate}
 \langle\mathcal{O}\rangle=\frac{4\xi}{\Delta_-}\,\lambda^{\frac{\Delta_+}{\Delta_-}}\,,
\end{equation}
where the coefficient of the $\phi^{4/\Delta_-}$ term in the superpotential is proportional to $\xi$.

Finally, the above discussion can also be framed in terms of the $\beta$-function~\cite{Megias:2014iwa} leading to $\Delta_\pm$-type solutions of the form
\begin{equation} \label{eq:UV_beta_function}
 \beta(\phi)=-\Delta_\pm\phi \,+\, \ldots\,,
\end{equation}
where once again the $\Delta_-$-type solutions admit a leading non-perturbative term of the form $\phi^{4/\Delta_--1}$.

%%%%%%%%%%%%%%%%%%%%%%%%%%%%%%%%%%%%%%%%%%%%%%%%%%%%%%%%%%%%%%%%%%%%%%%%%%%%%%%%%%%%%%%%%%%%%%%%%%%
\section{An Analytic Superpotential Model} \label{sec:analytic_solution}

After this brief introduction we next present simple examples that exhibit the possible properties of the dilaton. We shall initially consider the following simple model described by the superpotential
\begin{equation} \label{eq:superpotential}
 W(\phi)=3kM^3 \cosh\left(\sqrt{\frac{\alpha}{3M^3}}\,\phi\right),
\end{equation}
where $\alpha$ is a constant parameter.
This superpotential is analytic in $\phi^2$ and will allow us to straightforwardly solve for the metric and scalar backgrounds. As we shall see, the above choice of superpotential leads to the 5D gravity dual of a strongly coupled conformal sector where conformal invariance is spontaneously broken via the condensation of a finite dimensional operator. Also note that by defining our model in terms of the superpotential we have implicitly fixed the integration constant in Eq.~\eqref{eq:superpotential_def}.

\begin{figure}[h]
\begin{minipage}[b]{0.5\linewidth}
\begin{center}
\includegraphics[height=5.5cm]{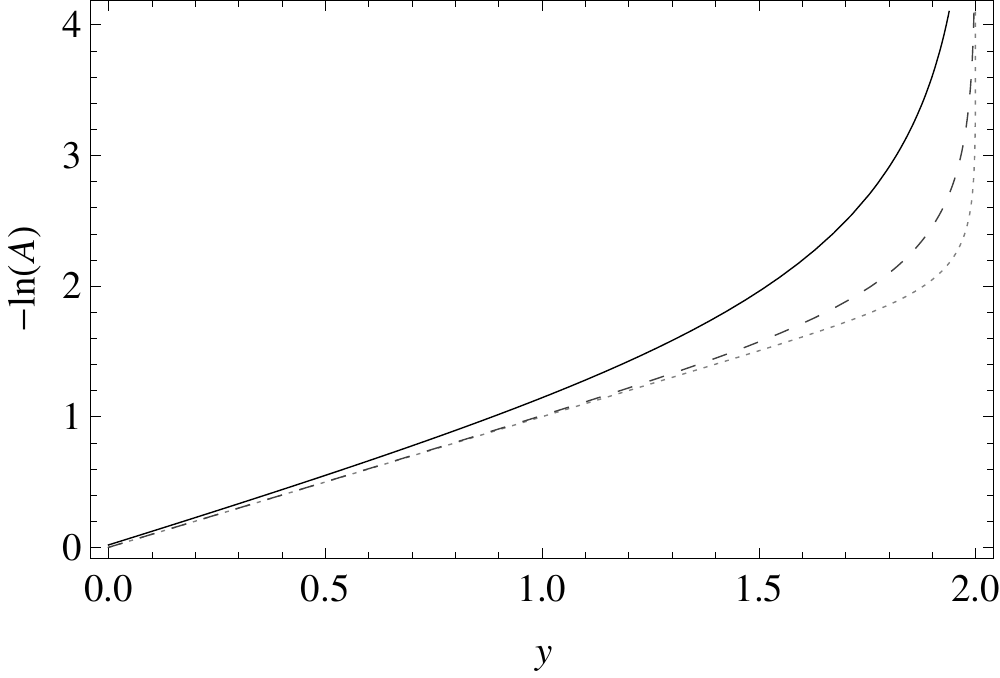}
\end{center}
\end{minipage}
\begin{minipage}[b]{0.5\linewidth}
\begin{center}
\includegraphics[height=5.5cm]{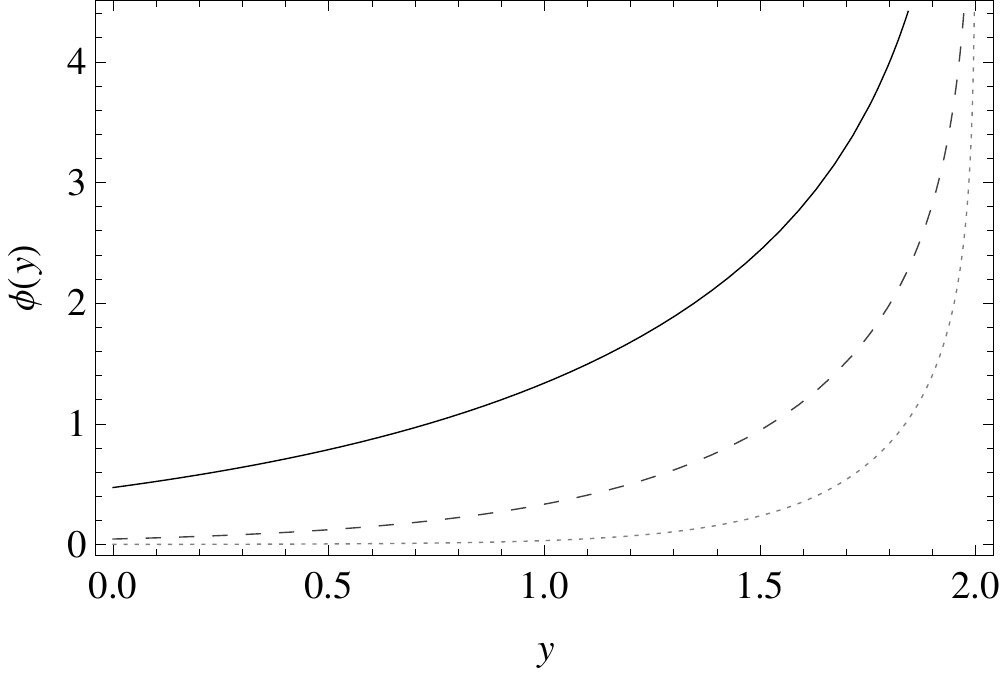}
\end{center}
\end{minipage}
\caption{Background solutions for the warp factor and scalar profile as a function of $y$. We have set $M=1$, $k=1$, and $y_c=2$. The solid, dashed and dotted curves correspond to $\alpha=1,2$ and 4 respectively.}
\label{fig:bkg_solutions}
\end{figure}

The bulk potential corresponding to our superpotential is
\begin{equation} \label{eq:bulk_potential}
 V(\phi)=\frac{3}{2}k^2M^3\left[(\alpha-4)\cosh^2\left(\sqrt{\frac{\alpha}{3M^3}}\,\phi\right)-\alpha\right]\,.
\end{equation}
Notice that this reduces to the familiar constant potential $V=-6k^2M^3$ if we take $\alpha=4$\footnote{This case corresponds to the model considered in \cite{Bellazzini:2013fga}. As we shall discuss, values of $\alpha\geq4$ require the introduction a hard cut-off in the IR. By considering the more general form for the superpotential \eqref{eq:superpotential} we are able to construct a true soft-wall model of spontaneous conformal symmetry breaking.}. Solving the equation of motion \eqref{eq:bulk_bkg_eom} we obtain the background solution
\begin{align} \label{eq:bkg_solution}
 A(y)&=e^{-ky_c}\bigg[2\sinh(k\alpha(y_c-y))\bigg]^{\frac{1}{\alpha}}, \notag \\ 
 \phi(y)&=-\sqrt{\frac{3M^3}{\alpha}}\log\left[\tanh\left(\frac{1}{2}k\alpha(y_c-y)\right)\right].
\end{align}
This solution has an AdS vacuum as $y_0\rightarrow-\infty$, which we have chosen to be at $\phi=0$ without loss of generality. Notice that the scalar profile is singular in the IR and there is a naked curvature singularity at $y=y_c$. We therefore have a dynamically generated boundary or ``soft-wall" with spacetime ending at $y=y_c$. The solutions are plotted in Figure~\ref*{fig:bkg_solutions}.

For completeness we also take the following UV boundary potential
\begin{equation}
 U(\phi)=\Lambda_{UV}+m_{UV}k(\phi-\phi_0)^2,
\end{equation}
where $\Lambda_{UV}$ is the UV brane tension and $m_{UV}$ is the UV boundary mass. However in most cases we will take the infinite boundary mass limit, which forces $\phi=\phi_0$ on the brane; the exact form of the potential is then unimportant. Imposing the boundary conditions \eqref{eq:bdry_bkg_eom} enforces the usual tuning of the 4D cosmological constant in addition to fixing the IR scale according to
\begin{equation} \label{eq:stabilisation}
 k(y_c-y_0)=\frac{2}{\alpha}\arctanh\left(e^{-\sqrt{\frac{\alpha}{3M^3}}\,\phi_0}\right),
\end{equation}
where we have taken the limit $m_{UV}\rightarrow\infty$. It is clear that generating a large hierarchy then requires a hierarchically small value of $\phi_0$. In the case of purely spontaneous breaking of the conformal symmetry this small boundary value for $\phi$ will be directly related to the ratio of the condensate and UV cut-off scales.

It is straightforward to evaluate the $\beta$-function corresponding to our superpotential to obtain
\begin{equation} \label{eq:beta}
 \beta(\phi)=-\sqrt{3M^3\alpha}\,\tanh\left(\sqrt{\frac{\alpha}{3M^3}}\,\phi\right),
\end{equation}
or expressed in terms of the background solutions
\begin{equation} \label{eq:beta_y}
 \beta(\phi(y))=\frac{A(y)\phi'(y)}{A'(y)}=-\sqrt{3M^3\alpha}\,\sech(k\alpha(y_c-y)).
\end{equation}
The limit on the asymptotic value of the $\beta$-function in Eq.~\eqref{eq:beta_limit} then leads to the requirement that
\begin{equation}
 \qquad1<\alpha<4.
\end{equation}
For $\alpha\leq1$ a continuous spectrum is obtained, while for $\alpha\geq4$ it becomes necessary to introduce a hard cut-off/IR brane in order to satisfy the boundary condition. In other words, our choice of superpotential no longer satisfies the IR boundary condition for $\alpha\geq4$. In the dual 4D description the introduction of an IR brane corresponds to the spontaneous breaking of the conformal symmetry by an infinite dimensional operator.

\begin{figure}[h]
\begin{minipage}[b]{0.5\linewidth}
\begin{center}
\includegraphics[height=5.5cm]{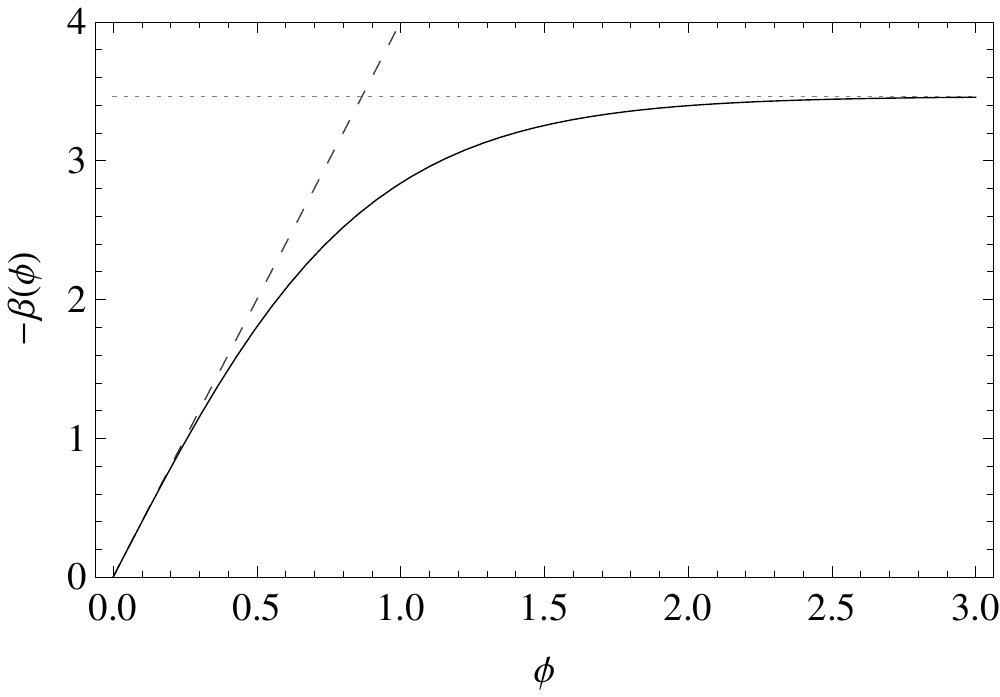}
\end{center}
\end{minipage}
\begin{minipage}[b]{0.5\linewidth}
\begin{center}
\includegraphics[height=5.35cm]{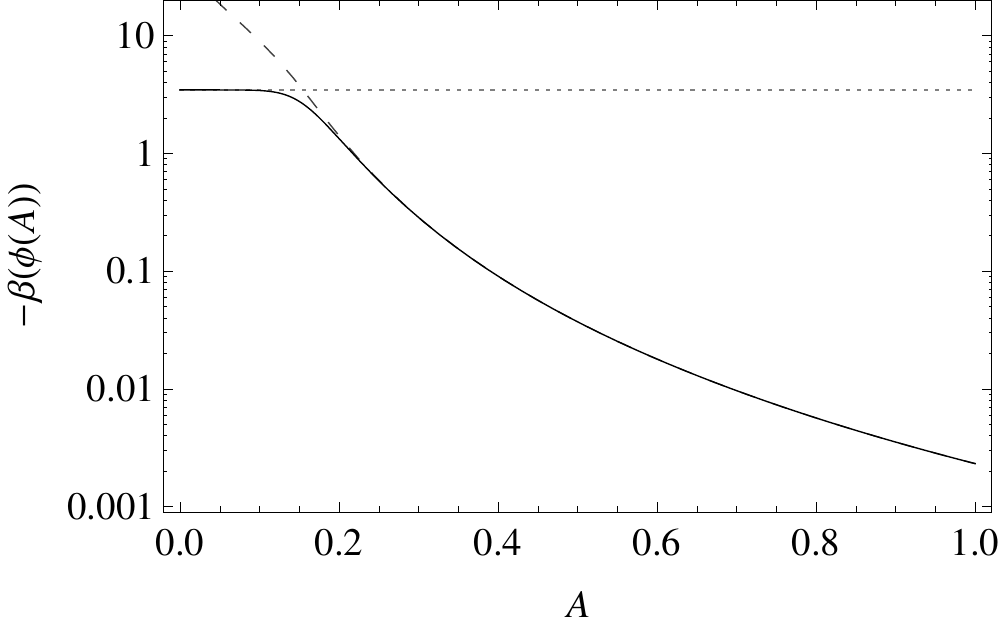}
\end{center}
\end{minipage}
\caption{Holographic $\beta$-function plotted as a function of $\phi$ and $A$ in the left and right panels respectively. The solid line corresponds to the full $\beta$-function~\eqref{eq:beta}, while the dashed line shows the leading-order term when expanded around $\phi=0$. The dotted line denotes the asymptotic value in the IR. We have set $M=1$, $k=1$, $\alpha=2$ and $y_c=2$.}
\label{fig:beta}
\end{figure}

We now want to look at the behaviour near the UV fixed point in order to interpret the soft-wall solution in terms of the dual CFT. Firstly, expanding the bulk potential~\eqref{eq:bulk_potential} around $\phi=0$ gives
\begin{equation} \label{eq:approx_bulk_potential}
 V(\phi)=-6k^2M^3+\frac{1}{2}k^2\alpha(\alpha-4)\phi^2+O(\phi^4)\,.
\end{equation}
This allows us to evaluate $\Delta_\pm$ for our solution according to
\begin{equation} \label{eq:delta}
\Delta_\pm=2\pm\sqrt{4+m_\phi^2/k^2}=2\pm\vert\alpha-2\vert.
\end{equation}
We can also expand~\eqref{eq:beta} around $\phi=0$ to obtain
\begin{equation} \label{eq:approx_beta}
 \beta(\phi)=-\alpha\,\phi\,+\,\ldots\quad=\left\{     
  \begin{array}{lr}
       -\Delta_-\phi\,+\,\ldots & , \quad\alpha<2\\
       -\Delta_+\phi\,+\,\ldots & , \quad\alpha\geq2
  \end{array}
 \right.\,.
\end{equation}
We therefore have a $\Delta_+$-type $\beta$-function for $\alpha\geq2$ which (for Dirichlet boundary conditions) corresponds to a CFT with purely spontaneous conformal symmetry breaking via the condensation of an operator $\mathcal{O}$ (i.e. $\lambda=0$). On the other hand for $\alpha<2$ our solution is of the $\Delta_-$-type and we have a deformed CFT where conformal invariance is explicitly broken by the coupling $\lambda$. Note that due to the fact that our superpotential is analytic in $\phi^2$ there can be no non-perturbative piece in the $\beta$-function ($\xi=0$) and therefore there is no condensate in this case. Depending on the value of $\alpha$, our solution therefore describes a CFT either spontaneously broken or deformed by an operator of dimension $2\leq[\mathcal{O}]<4$. If we instead consider the alternative CFT description and assume Neumann boundary conditions for the scalar then the situation is reversed and we have spontaneous breaking for $1<\alpha<2$ and a deformed CFT with no condensate 
for $2\leq\alpha\leq3$, corresponding to operator dimensions $1\leq[\mathcal{O}]\leq2$. The beta function is shown in Figure~\ref*{fig:beta}. 

%%%%%%%%%%%%%%%%%%%%%%%%%%%%%%%%%%%%%%%%%%%%%%%%%%%%%%%%%%%%%%%%%%%%%%%%%%%%%%%%%%%%%%%%%%%%%%%%%%%
\subsection{Fine-tuning and Spontaneous Breaking} \label{sec:tuning}

It is well known that a purely spontaneously broken CFT is expected to be tuned. The issue arises from the fact that unlike non-linearly realised internal symmetries, a non-derivative quartic coupling for the dilaton is permitted by scale invariance. Naive dimensional analysis suggests that generically this coupling should be large ($\sim16\pi^2$), however a massless dilaton arising from spontaneous breaking can only be obtained in the special case where the quartic coupling is zero. In the absence of some mechanism such as supersymmetry, which can give rise to flat directions in the dilaton potential, spontaneous breaking therefore requires a tuning of the quartic coupling. 

From the 5D viewpoint, this tuning can be readily identified in RS+GW type models as the tuning of the IR brane tension~\cite{Bellazzini:2012vz}; however in soft-wall models the situation is more subtle and related to the various superpotentials which can arise for a given bulk potential. Generically we expect a $\Delta_-$-type superpotential that is a non-analytic function of $\phi$. In addition there may be particular values of the integration constant in Eq.~\eqref{eq:superpotential_def} which lead to $\Delta_+$-type solutions or $\Delta_-$-type solutions with an analytic superpotential (i.e. $\xi=0$). Situations where such solutions not only exist but also satisfy the regularity conditions in the IR are clearly non-generic and generally arise from carefully constructed bulk potentials. In addition, by evaluating the on-shell action one can easily demonstrate that such solutions correspond to a dilaton potential where the quartic term is set to zero as we shall see in Section~\ref{sec:effective_potential}.

We note that while our solution should therefore be considered tuned, it nevertheless provides a useful model for a spontaneously broken conformal sector where the background equations can be solved analytically. In other words our solution corresponds to a soft-wall version of pure RS, but describes the more realistic case of spontaneous breaking by a finite-dimensional operator. In Section~\ref*{sec:general_case} we shall consider a more general bulk potential to investigate the more generic (un-tuned) case where we have a non-zero quartic term in the dilaton potential.

%%%%%%%%%%%%%%%%%%%%%%%%%%%%%%%%%%%%%%%%%%%%%%%%%%%%%%%%%%%%%%%%%%%%%%%%%%%%%%%%%%%%%%%%%%%%%%%%%%%
\subsection{Mass Spectrum} \label{sec:mass_spectrum}

Next we determine the mass spectrum for the scalar modes of the solution~\eqref{eq:bkg_solution}. In the case of massive modes, the metric and bulk scalar perturbations can be parameterised by
\begin{eqnarray} \label{eq:perturbations}
 g_{\mu\nu} &=& A^2(y)\left[\eta_{\mu\nu}(1+2\Psi)+2\partial_\mu\partial_\nu E+2\partial_{(\mu}E^T_{\nu)}+h_{\mu\nu}^{TT}\right],\\
  g_{\mu 5} &=& A(y)\left[\partial_\mu B+B_\mu^T\right],\\
  g_{55} &=& 1+2\Phi,\\
  \phi &=& \phi(y)+\delta\phi(x,y),
\end{eqnarray}
where $\partial^\mu B_\mu^T=\partial^\mu E_\mu^T=\partial^\mu h_{\mu\nu}^{TT}=h^{TT\mu}_\mu=0$. After performing a Kaluza-Klein (KK) expansion, the dynamical equation for the scalar sector can be written as~\cite{Kofman:2004tk}
\begin{equation} \label{eq:scalar_eom}
 \frac{1}{A^4\beta^2}\frac{d}{dy}\bigg[A^4\beta^2\zeta_n'\bigg]+\frac{m_n^2}{A^2}\zeta_n=0,
\end{equation}
where $\zeta=\Psi-\delta\phi/\beta$ is a gauge invariant variable. There is also the constraint equation
\begin{equation} \label{eq:constraint}
 \Psi'-\frac{A'}{A}\Phi+\frac{1}{3M^3}\phi'\,\delta\phi=0.
\end{equation}
The UV boundary condition is given by
\begin{equation} \label{eq:scalar_UV_bc}
 \delta\phi'-\phi'\Phi=\frac{1}{2}U''(\phi)\,\delta\phi,
\end{equation}
while at the dynamical IR boundary we have
\begin{equation} \label{eq:scalar_IR_bc}
 A^4\delta\phi'=0.
\end{equation}
There will also be an additional normalisation condition.

%%%%%%%%%%%%%%%%%%%%%%%%%%%%%%%%%%%%%%%%%%%%%%%%%%%%%%%%%%%%%%%%%%%%%%%%%%%%%%%%%%%%%%%%%%%%%%%%%%%
\subsubsection{Massless dilaton from spontaneous breaking} 

We begin by investigating under which circumstances it is possible to obtain a massless dilaton from spontaneous breaking of conformal invariance. Note however that for massless modes the above decomposition~\eqref{eq:perturbations} becomes ambiguous and the metric perturbations should instead be decomposed in terms of a light-cone basis. The solution then contains two dynamical scalar degrees of freedom $c_{1,2}$ and is given after gauge fixing by~\cite{kiritsis:2006ua}
\begin{equation}
 \zeta_1=c_1(x),\qquad \zeta_2=\frac{A'}{A^3}\,c_2(x)-2\left(\frac{A'}{A^3}\int dy\,A^2\right)(y)\,c_1(x),
\end{equation}
where $\zeta_1=\Psi-\delta\phi/\beta$ and $\zeta_2=\delta\phi/\beta$. We can immediately remove the $c_2$ mode by imposing the IR boundary condition, leaving just a single 4D massless mode, $c_1$.

In the presence of a finite UV cut-off it is expected that the dilaton will acquire a mass due to the coupling of the CFT to the dynamical source. This indeed turns out to be the case and can be easily checked by imposing the UV boundary condition~\eqref{eq:scalar_UV_bc} for finite $y_0$. One then finds that the above solution does not satisfy the boundary conditions, with the usual exception when there is a tuning between the boundary and bulk masses. On the other hand, as the UV cut-off is removed ($y_0\rightarrow-\infty$) the source decouples and we should expect a massless dilaton if the conformal symmetry is spontaneously broken. Imposing Dirichlet or Neumann boundary conditions for $\delta\phi$ requires that the leading ($A^{\Delta_-}$) or subleading ($A^{\Delta_+}$) terms vanish respectively. We then find that the remaining mode, $c_1$, does indeed satisfy the boundary conditions for $\alpha\geq2$ (Dirichlet) or $\alpha<2$ (Neumann) as expected.

%%%%%%%%%%%%%%%%%%%%%%%%%%%%%%%%%%%%%%%%%%%%%%%%%%%%%%%%%%%%%%%%%%%%%%%%%%%%%%%%%%%%%%%%%%%%%%%%%%
\subsubsection{Light dilaton with UV explicit breaking}

Moving now to the case of a massive dilaton, the gauge freedom allows us to make the choice $\Psi=0$ and express Eq.~\eqref{eq:scalar_eom} solely in terms of $\delta\phi$. Using \eqref{eq:constraint} the UV boundary condition can similarly be written in terms of $\delta\phi$ as
\begin{equation}
 \delta\phi'=\left(\frac{1}{2}U''(\phi)+\frac{1}{3M^3}\phi'\beta\right)\delta\phi.
\end{equation}

While the equation of motion is non-trivial and can in general only be solved numerically, an approximate solution for $m\ll m_{KK}$ can be found by matching solutions in the UV and IR, where $m_{KK}=[\int dy/A]^{-1}k$. Generalising the procedure used in~\cite{Megias:2014iwa} to allow for general boundary conditions we obtain the following expression for the dilaton mass, $m_D$
\begin{equation} \label{eq:approx_dil_mass}
 \frac{1}{m_{D}^2} \simeq \left(\int^{y_c}_{y_0}dy\,\frac{1}{A^4\beta^2}\int^{y_c}_y dy'\,A^2\beta^2 \right)+ \kappa \int^{y_c}_{y_0} dy\,A^2\beta^2,
\end{equation}
where
\begin{equation} \label{eq:gamma}
 \kappa=\frac{1}{A^4\beta^2}\left(\frac{1}{2}U''(\phi)+\frac{1}{3M^3}\phi'\beta-\frac{\beta'}{\beta}\right)^{-1}\bigg\vert_{y_0},
\end{equation}
and $\beta'$ in (\ref{eq:gamma}) is the derivative w.r.t $y$.
Unfortunately in our case the  first term above can still only be evaluated numerically\footnote{A reasonable approximation can be obtained by noticing that the inner integral gives a constant of order $m_{KK}^2$ for all values of $y$, except near $y=y_c$. An analytic expression for the dilaton mass can then be obtained by evaluating the outer integral and keeping only the dominant UV boundary term.}. Taking the $m_{UV}\rightarrow\infty$ limit, which forces $\delta\phi=0$ on the UV boundary, we have $\kappa\rightarrow 0$ and the expression reduces to that given in Ref.~\cite{Megias:2014iwa}. 

\begin{figure}[h]
\begin{minipage}[b]{0.5\linewidth}
\begin{center}
\includegraphics[height=5cm]{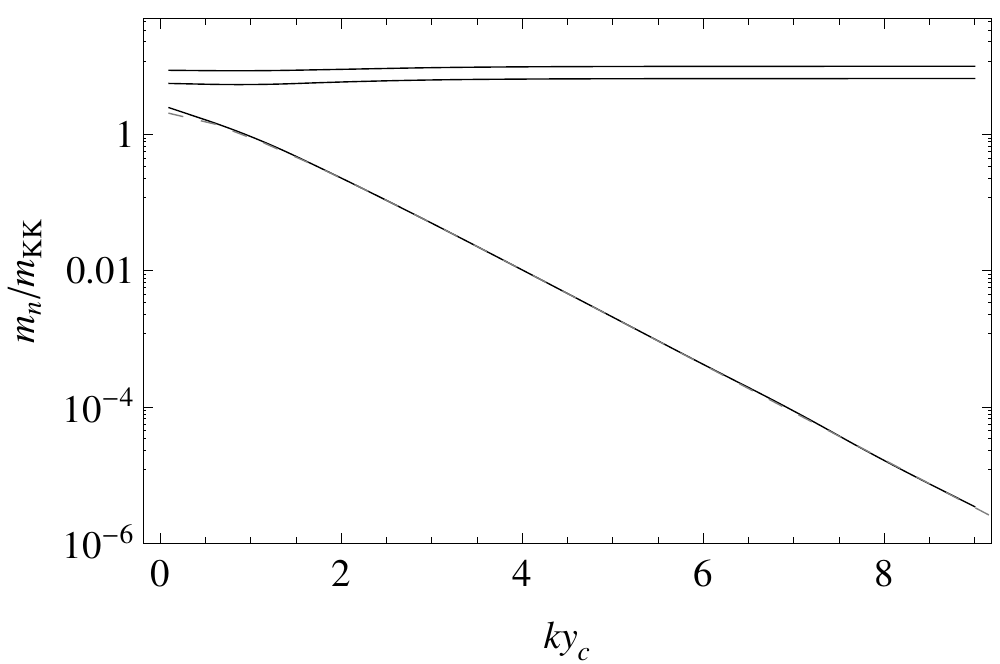}
\end{center}
\end{minipage}
\begin{minipage}[b]{0.5\linewidth}
\begin{center}
\includegraphics[height=5cm]{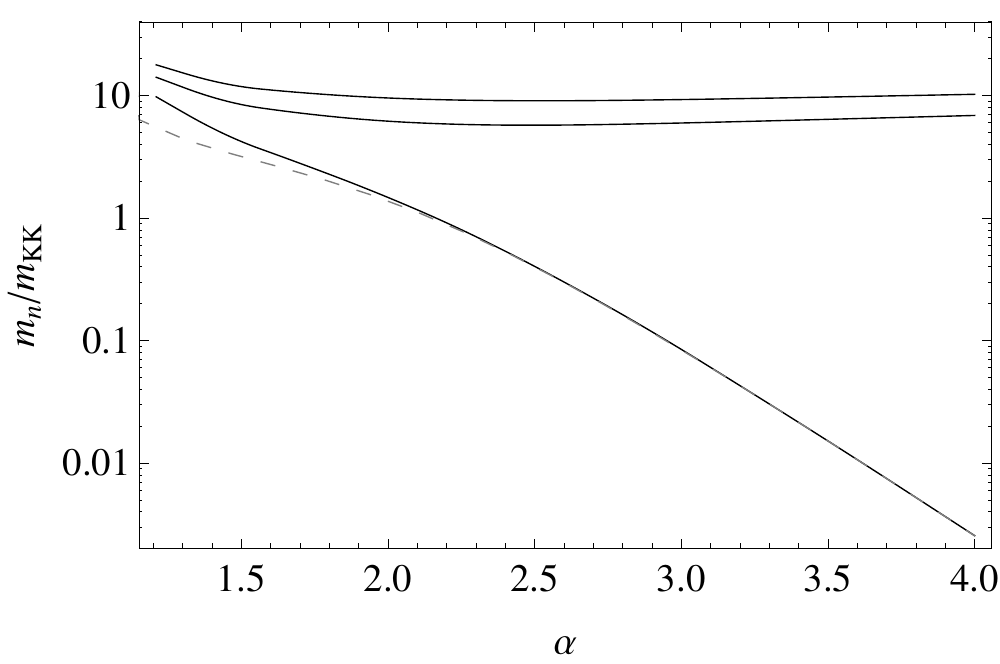}
\end{center}
\end{minipage}
\caption{Masses of the three lightest modes as a function of $ky_c$ for $\alpha=3.6$ (left panel) and as a function of $\alpha$ for $ky_c=4$ (right panel). The dashed line shows the approximate solution~\eqref{eq:approx_dil_mass}. We have set $y_0=0$.}
\label{fig:dilaton_mass}
\end{figure}

We have also solved Eq.~\eqref{eq:scalar_eom} numerically for the masses and profiles of the dilaton and the lowest few KK modes using the shooting method. The results are given in Figure~\ref*{fig:dilaton_mass}, where the dashed line shows the approximate solution~\eqref{eq:approx_dil_mass} and we have taken $m_{UV}\rightarrow\infty$. Note that the approximate solution is in good agreement with the full numerical solution in the cases where we have a light dilaton. 

Focusing initially on the left panel, where $\alpha=3.6$, we find that there exists a very light mode as the separation between the UV brane and the soft-wall becomes large. This is not surprising, since for $\alpha\geq2$ we expect to obtain a massless mode from the spontaneous breaking when the explicit breaking effects of the UV cut-off are decoupled. In this scenario the dilaton mass is arising solely due to the coupling of the CFT to the dynamical source and the small dilaton mass is directly related to the fact that $\mu_0\gg m_{KK}$, where the UV scale $\mu_0\equiv k A(y_0)$. A similar effect for a soft-wall background was also obtained in Ref.~\cite{Gherghetta:2011rr}. For large hierarchies it is therefore possible to exponentially suppress the dilaton mass relative to the confinement/symmetry breaking scale. We emphasise however that the presence of such a light mode is dependent on the fact that we have a $\Delta_+$-type solution in this regime, and as discussed in Section~\ref*{sec:beta_function} 
such scenarios should be considered tuned. Finally, it is worth noting that this contribution to the dilaton mass does not appear in hard-wall models of spontaneous breaking, where the dilaton is a purely composite state. In the standard RS case the dilaton remains massless even in the presence of the UV brane due to the fact that the IR brane corresponds to spontaneous breaking by an infinite dimensional operator and is therefore not sensitive to the UV cut-off.

\begin{figure}[h]
\begin{center}
\includegraphics[height=7cm]{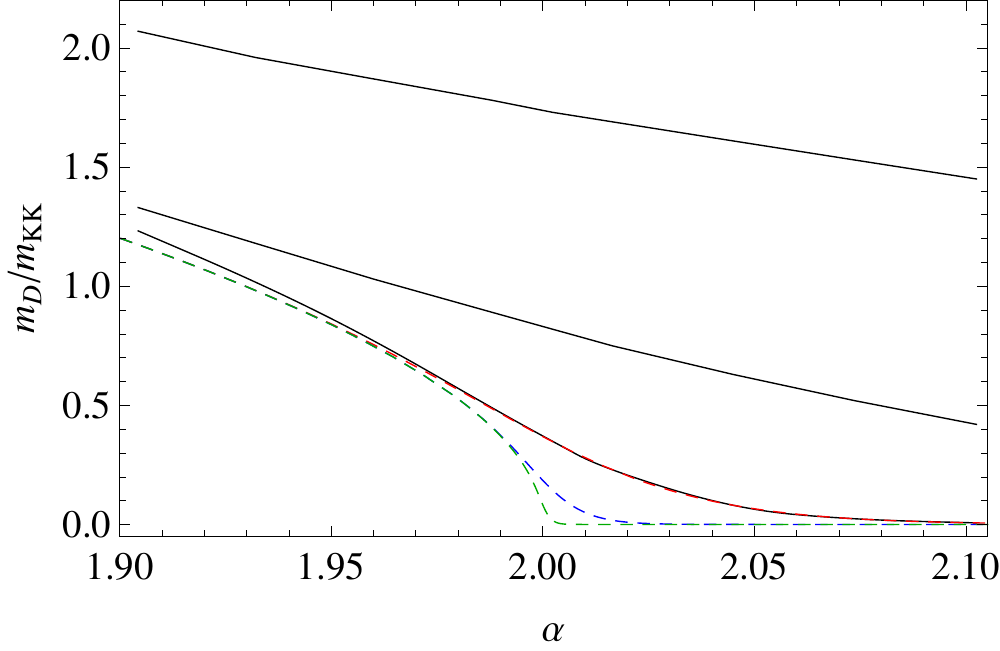}
\end{center}
\caption{Dilaton mass near $\alpha=2$ for $ky_0=-2$, $-10$, $-50$, $-200$, $-1000$ (top to bottom). The dashed curves are obtained using Eq.~\eqref{eq:approx_dil_mass}. We have set $ky_c=1$.}
\label{fig:dilaton_mass_alpha_2}
\end{figure}

Considering now the right panel in Figure~\ref{fig:dilaton_mass}, we see that for smaller values of $\alpha$ (in particular $\alpha<2$) there is no longer a light mode in the spectrum. This is again to be expected, since in this regime we have a $\Delta_-$-type solution corresponding to a deformed CFT. The dilaton therefore acquires a large mass even in the absence of a UV cut-off. This is shown more clearly in Figure~\ref*{fig:dilaton_mass_alpha_2}, where we have plotted the dilaton mass near $\alpha=2$ for various values of $y_0$. As discussed previously, for $\alpha>2$ we obtain a massless mode as $y_0\rightarrow-\infty$, while for $\alpha<2$ the dilaton quickly becomes massive. 

Interestingly it seems possible to obtain a light dilaton from a deformed CFT with a $\Delta_-$-type solution when approaching $\alpha=2$ from below, albeit only for large hierarchies. The existence of this light mode can be understood by looking at the behaviour of Eq.~\eqref{eq:approx_dil_mass} in the limit $y_0\rightarrow-\infty$. Firstly, the inner integrand in this expression simply gives a constant of order $m_{KK}^2$ for all values of $y$ (except near $y=y_c$) and so the behaviour of the mass is largely determined by the outer integral. Now, in the UV the metric is approximately AdS and we can rewrite this as an integral over the warp factor, $A$. Then using the fact that the $\beta$-function in the UV is well approximated by $\beta(\phi)=-\Delta_-\phi$ and substituting in the leading behaviour $\phi\sim A^{-\Delta_-}$ we obtain
\begin{equation}
\int dA\,A^{2\Delta_--5}\,.
\end{equation}
For $\Delta_-=2$ this integral clearly becomes divergent as $A\rightarrow\infty$ ($y_0\rightarrow-\infty$), giving rise to a massless dilaton. Near $\Delta_-\lesssim2$ the dilaton mass is then described by
\begin{equation}
 m_D\sim\sqrt{2-\Delta_-}\, m_{KK}\,.
\end{equation}
We therefore find a light dilaton with an operator of dimension 2, in contrast to the well-studied scenario where a naturally light dilaton is obtained with a near-marginal operator~\cite{Bellazzini:2013fga,Coradeschi:2013gda}. However, the presence of this light mode is a consequence of the particular form of our bulk potential and relies upon the fact that our solution transitions to a $\Delta_+$-type solution at $\alpha=2$. The light mode is therefore a result of the tuning in this model arising from the absence of a non-perturbative term in the superpotential. For a more general bulk potential, where the solution remains of the $\Delta_-$-type, the leading order behaviour of $\phi$ at $\Delta=2$ is given by $\phi\sim\log(A)A^{-2}$. This logarithmic behaviour ensures that the above integral remains finite and there is no light mode.

%%%%%%%%%%%%%%%%%%%%%%%%%%%%%%%%%%%%%%%%%%%%%%%%%%%%%%%%%%%%%%%%%%%%%%%%%%%%%%%%%%%%%%%%%%%%%%%%%%%
\subsection{Effective Potential} \label{sec:effective_potential}

Similar to the approach commonly undertaken in hard-wall models~\cite{Goldberger:1999uk,Bellazzini:2012vz}, it is useful to construct a 4D effective potential for the dilaton by integrating out the extra dimension. After using the bulk equations of motion, it is straightforward to show that the effective potential in the presence of a soft-wall is given by
\begin{equation} \label{eq:eff_potential}
 V_{eff}=A^4(y_0)\,\left[U(\phi(y_0))-2W(\phi(y_0))\right]-\frac{2}{3}\lim_{y\to y_c}{A^4(y)\,W(\phi(y))}.
\end{equation}
Provided that the superpotential satisfies certain boundary conditions in the IR, as discussed in Section~\ref{sec:beta_function}, the last term in the above expression will be zero and can be ignored. Once again this will be the case for our solution provided $\alpha<4$. 

\begin{figure}[h]
\begin{minipage}[b]{0.5\linewidth}
\begin{center}
\includegraphics[height=5.5cm]{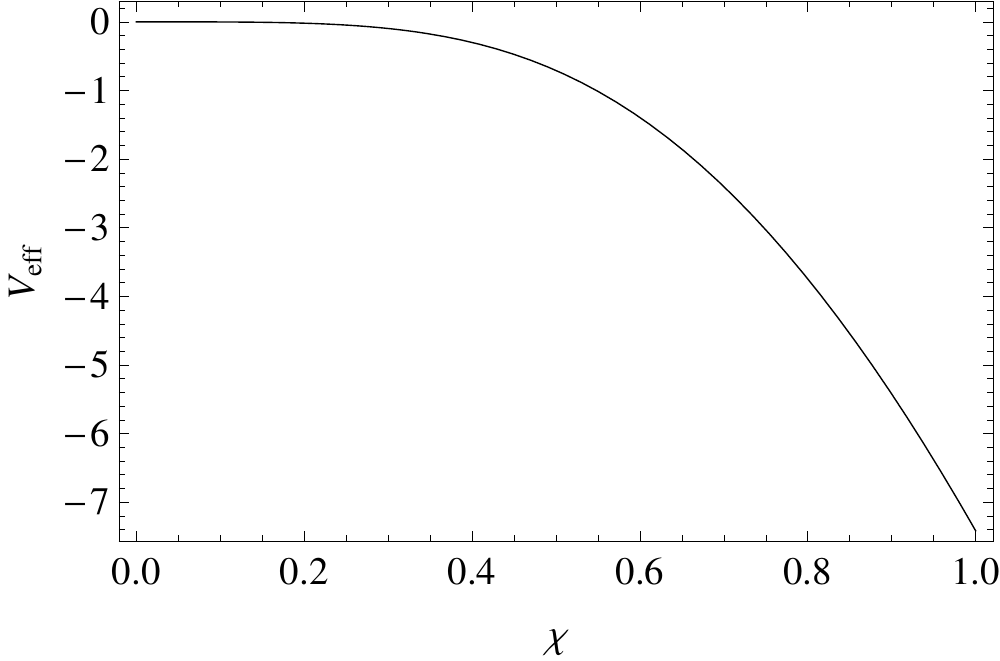}
\end{center}
\end{minipage}
\begin{minipage}[b]{0.5\linewidth}
\begin{center}
\includegraphics[height=5.5cm]{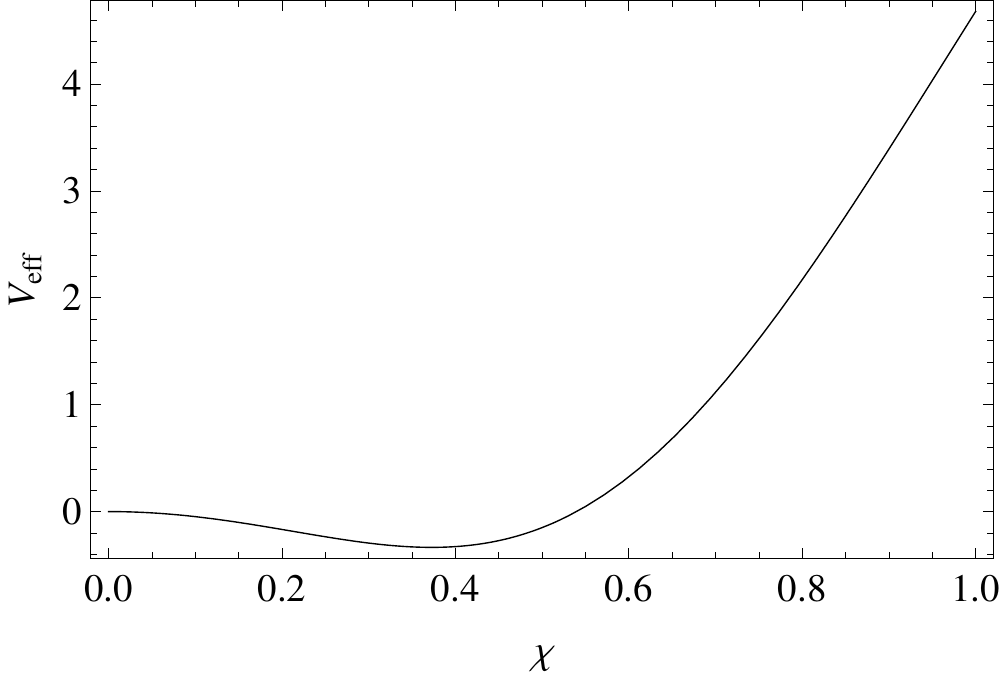}
\end{center}
\end{minipage}
\caption{Effective potential, where the dilaton has been parameterised by $\chi=e^{-ky_c}$. The left panel corresponds to the case $m_{UV}=0$, while the right panel has $m_{UV}=5$ and $\phi_0=0.2$. We have set $M=k=\mu_0=1$, $\alpha=2$ and fixed $\Lambda_{UV}=6kM^3-k\,m_{UV}\phi_0^2$.}
\label{fig:Veff}
\end{figure}

It is worth pointing out that unlike the hard-wall case the warp-factor depends on the dilaton vacuum expectation value through $y_c$. Therefore simply minimising Eq.~\eqref{eq:eff_potential} for fixed $y_0$ may lead to unphysical minima which do not satisfy the boundary equations of motion \eqref{eq:bdry_bkg_eom}. Instead one should minimise the potential while keeping the physical UV scale $\mu_0$ fixed. Then, parameterising the dilaton by\footnote{One could also choose to parameterise the dilaton by the inverse conformal volume, $(kz_c)^{-1}$. However it can be easily shown that the two parameterisations are equivalent up to a simple rescaling, which depends only on $\alpha$.} $\chi\equiv e^{-ky_c}$, the effective potential is given by
\begin{equation} \label{eq:chi_potential}
 \frac{V_{eff}(\chi)}{\mu_0^4}=\Lambda_{UV} + m_{UV}k\left(\phi_0+\sqrt{\frac{3M^3}{\alpha}}\,\log(F(\chi))\right)^2-3kM^3\left(F(\chi)+\frac{1}{F(\chi)}\right),
\end{equation}
where 
\begin{equation}
 F(\chi)=\left(\frac{\chi}{\mu_0}\right)^\alpha\left(-2+\sqrt{4+\left(\frac{\chi}{\mu_0}\right)^{-2\alpha}}\right).
\end{equation}
The effective potential is plotted in Figure~\ref*{fig:Veff}. The left panel shows the unstabilised case in the absence of a boundary mass term, while in the right panel we can clearly see the generation of a non-zero minimum for the dilaton once the bulk scalar obtains a vacuum expectation value on the boundary. 

While the general expression~\eqref{eq:chi_potential} is complicated, the situation becomes much simpler as the UV cut-off is removed. Taking the limit $\mu_0\rightarrow\infty$ and adding appropriate counterterms to subtract the divergent terms\footnote{The specific counterterms necessary to subtract the divergent terms in Eq.~\eqref{eq:chi_potential} are given by $6kM^3+3k\frac{M^3}{\alpha}\Delta_-\,\log(F(\chi))^2$.} leads to
\begin{equation} \label{eq:eff_pot_inf}
 V_{eff}(\chi)=\left\{ 
  \begin{array}{lr}
   {\bar\Lambda_{UV}} + k\, \bar{m}_{UV}\bar{\phi}_0^2 - 4k \sqrt{\frac{3M^3}{\alpha}} \bar{\phi}_0 \bar{m}_{UV}\,\chi^{\alpha} + 12k \frac{M^3}{\alpha}\bar{m}_{UV}\,\chi^{2\alpha} & , \quad\alpha<2\\
   {\bar\Lambda_{UV}} + k\, \bar{m}_{UV}\bar{\phi}_0^2 & , \quad\alpha\geq2,
  \end{array}
 \right.
\end{equation}
where we have simultaneously performed the rescaling $\Lambda_{UV}={\bar\Lambda_{UV}}\,\mu_0^{-4}$, $\phi_0=\bar{\phi}_0\,\mu_0^{-\Delta_-}$ and $m_{UV}=\bar{m}_{UV}\,\mu_0^{2(\Delta_--2)}$. From the 4D viewpoint this rescaling simply corresponds to running the couplings up to the scale $\mu_0$ and ensures that the IR scale remains fixed when taking the limit $\mu_0\rightarrow\infty$. 

If we consider the alternative CFT description, where the operator dimension is given by $\Delta_-$, then the on-shell action is identified with the 1PI effective action in the CFT~\cite{Klebanov:1999tb}. We can then straightforwardly interpret Eq.~\eqref{eq:eff_pot_inf} as it is directly related to the effective potential in the alternative CFT. This becomes more evident if we compare Eqs.~\eqref{eq:UV_scalar_vev}\footnote{In this case the leading term in Eq.~\eqref{eq:UV_scalar_vev} is identified with the condensate, $\langle\mathcal{O}\rangle$.} and \eqref{eq:bkg_solution} to obtain the relation $\langle\mathcal{O}\rangle\sim\chi^\alpha$. 

Let us initially focus on $\alpha<2$ where we have a $\Delta_-$-type solution. If we then take $\bar{m}_{UV}=0$ in Eq.~\eqref{eq:eff_pot_inf} we obtain a flat potential, consistent with spontaneous breaking by an operator of dimension $\Delta_-$. It is worth noting however that in this case the potential becomes unstable for finite $\mu_0$. For non-zero $\bar{m}_{UV}$, the boundary conditions are modified and the two $\chi$-dependent terms in Eq.~\eqref{eq:eff_pot_inf} can be identified with turning on a source for a single and double-trace deformation in the CFT~\cite{Witten:2001ua}. This results in a non-zero condensate and minimising \eqref{eq:eff_pot_inf} for $\alpha<2$ we obtain 
\begin{equation}
 \langle\chi\rangle=\left(\frac{1}{2}\sqrt{\frac{\alpha}{3M^3}}\,\bar{\phi}_0\right)^{\frac{1}{\alpha}}.
\end{equation}
It can also be shown that this is consistent with the result obtained by directly solving the boundary conditions by taking the limit $y_0\rightarrow\-\infty$ in Eq.~\eqref{eq:stabilisation}. 

We now turn to the case $\alpha\geq2$ where we have a $\Delta_+$-type solution. When considering the alternative CFT description, $\Delta_+$-type solutions can only describe the situation where $\langle\mathcal{O}\rangle=0$ (because the leading term is absent in \eqref{eq:UV_scalar_vev}) and so a non-trivial potential cannot be generated in Eq.~\eqref{eq:eff_pot_inf}. Since there is no spontaneous breaking we do not obtain a massless mode with Neumann boundary conditions when $\alpha\geq2$ as seen in Section~\ref{sec:mass_spectrum}. If we move to the standard CFT description instead, $\Delta_+$-type solutions now describe the case of zero source and this means that Eq.~\eqref{eq:eff_pot_inf} can still be interpreted as the effective potential in the CFT. We then have a flat potential consistent with spontaneous breaking by an operator of dimension $\Delta_+$.

Importantly, in the case of both $\Delta_-$ and $\Delta_+$-type solutions, the only terms which survive the $\mu_0\rightarrow\infty$ limit in Eq.~\eqref{eq:eff_pot_inf} are boundary terms and there is generally no quartic term in the potential. As mentioned in Section~\ref{sec:tuning} this is due to the fact that we have chosen a superpotential which is analytic in $\phi^2$. More generally the superpotential will include a leading non-analytic term of the form $\phi^{4/\Delta_-}$, which directly gives rise to a non-zero quartic term in the effective potential. This can be easily shown using Eqs.~\eqref{eq:UV_scalar_vev} and \eqref{eq:condensate} and then identifying the dilaton according to $\langle\mathcal{O}\rangle\sim\chi^{\Delta_+}$ in the case of standard quantisation or $\langle\mathcal{O}\rangle\sim\chi^{\Delta_-}$ for the alternative quantisation.

Finally, it is worth highlighting that an accurate expression for the dilaton mass cannot generally be obtained by simply expanding the effective potential around the minimum. While this may provide a good approximation to the dilaton mass in certain regions, for example $\alpha>2$, it fails to capture the correct behaviour in other regions such as near $\alpha\lesssim2$. In order to obtain the physical dilaton mass one should of course compute the full effective action, in particular the dilaton kinetic term.

%%%%%%%%%%%%%%%%%%%%%%%%%%%%%%%%%%%%%%%%%%%%%%%%%%%%%%%%%%%%%%%%%%%%%%%%%%%%%%%%%%%%%%%%%%%%%%%%%%%
\section{An Analytic Potential Model} \label{sec:general_case}

So far we have been considering a simple model with an analytic superpotential, which allowed us to straightforwardly solve for the warp factor and bulk scalar profile. While this model provides a useful description of a conformal sector spontaneously broken by a finite dimensional operator, we have seen that it corresponds to the special case of a vanishing quartic term in the dilaton potential. In the absence of some symmetry (e.g. supersymmetry) to enforce this condition, we must consider such cases tuned and instead look to a more general model.

Simply considering a superpotential which is non-analytic in $\phi$ will generally not lead to a bulk potential with an AdS extremum and so one should instead begin by choosing a general form for the bulk potential and then solve Eq.~\eqref{eq:superpotential_def} to obtain the superpotential. Unfortunately the cases where there exists a known solution to \eqref{eq:superpotential_def} are limited and for most bulk potentials the superpotential can only be determined numerically. However, such a numerical approach is indeed still useful and will allow us to determine the background solutions and dilaton mass for more general bulk potentials.

The simplest extension of our previous superpotential model leads to the following form for the bulk potential
\begin{equation} \label{eq:general_potential1}
 V(\phi)=\frac{3}{2} k^2M^3\left[\tau\cosh^2\left(\sqrt{\frac{\alpha}{3M^3}}\,\phi\right)-(4+\tau)\right],
\end{equation}
which is parameterised by the two constant parameters $\alpha$ and $\tau$ and reduces to Eq.~\eqref{eq:bulk_potential} in the special case $\tau=\alpha-4$. The above potential will indeed alleviate the tuning found in our simple model and consequently give rise to a non-zero quartic term in the dilaton potential. However we have numerically verified there is no light dilaton in the spectrum except in the regions of parameter space where we approach the tuned solution. 

We will therefore focus instead on a different form for the bulk potential, which as we shall see can indeed give rise to a naturally light dilaton. We will be interested in the following potential
\begin{equation} \label{eq:general_potential}
 V(\phi)=V_0\,e^{\tanh\left(\frac{\gamma}{\nu}\right)\sqrt{\frac{\alpha}{3M^3}}\phi}\,\cosh^{\nu}\left[\frac{1}{\nu}\left(\gamma-\sqrt{\frac{\alpha}{3M^3}}\phi\right)\right],
\end{equation}
parameterised by the constant parameters $\alpha,\gamma, \nu$ and where 
\begin{equation}
 V_0=-6k^2M^3\sech^{\nu}\left(\frac{\gamma}{\nu}\right).
\end{equation}
The coefficient of $\phi$ in the exponent has been chosen such that the potential once again possesses an AdS extremum at $\phi=0$. Expanding around this point gives 
\begin{equation}
 V(\phi)=-6k^2M^3-\frac{\alpha}{\nu}\sech^2\left(\frac{\gamma}{\nu}\right)k^2\,\phi^2+O(\phi^3)\,,
\end{equation}
and we again identify 
\begin{equation} \label{eq:delta_pm}
 \Delta_\pm=2\pm\sqrt{4-\frac{2\alpha}{\nu}\sech^2\left(\frac{\gamma}{\nu}\right)}.
\end{equation}

This potential has a couple of interesting features. Firstly, in certain regions of parameter space (e.g.\,$\vert \nu\vert\ll1$) it possess an approximate shift symmetry, $\phi\rightarrow\phi+c$, in the UV. Furthermore, for large $\phi$ the potential behaves as a simple exponential and this shift symmetry is then realised as an overall rescaling. It was pointed out in Ref.~\cite{Megias:2014iwa} that such a potential may be considered technically natural provided that the intermediate symmetry breaking region is small. Interestingly they also demonstrated that such a condition should also be satisfied in order for the spectrum to contain a light dilaton. We shall see this explicitly for our potential below.

In order to obtain a solution which satisfies the IR boundary condition, it will be numerically simpler to solve for the $\beta$-function rather than directly solving Eq.~\eqref{eq:superpotential_def} for the superpotential. Combining \eqref{eq:superpotential_def} with the definition of the $\beta$-function \eqref{eq:beta_def} yields the following equation
\begin{equation} \label{eq:beta_function}
 \beta'\beta=\left(\beta^2-12M^3\right)\left(\frac{\beta}{3M^3}+\frac{V'}{2V}\right),
\end{equation}
which is an example of Abel's equation of the second kind. It is clear from the above equation that it is the quantity $V'/V$, rather than the potential itself, which determines the behaviour of the $\beta$-function. Let us then consider more carefully the following expression
\begin{equation} \label{eq:V'V}
 \frac{V'(\phi)}{V(\phi)}=\sqrt{\frac{\alpha}{3M^3}}\left(\tanh\left[\frac{\gamma}{\nu}\right]-\tanh\left[\frac{1}{\nu}\left(\gamma-\sqrt{\frac{\alpha}{3M^3}}\phi\right)\right]\right).
\end{equation}
We can now clearly identify the role of the various parameters in our potential. For $\nu\ll1$ the above expression behaves as a step-function, with the location of the inflection point given by $\phi=\sqrt{3M^3/\alpha}\,\gamma$. The steepness of the step is determined by $\alpha/\nu$, while for small $\nu$ the asymptotic value in the IR is largely determined by $\alpha$. There are then three distinct regions for $V'/V$: two approximately flat regions for small and large $\phi$, and an intermediate transition region. We shall find that the $\beta$-function also exhibits a very similar behaviour with the three regions corresponding to a $\beta$-function which is walking in the UV; has an intermediate region identified with the turning on of a condensate; and then confines in the IR.

Recall that for a $\beta$-function which is both confining and satisfies the IR boundary conditions, there is limit on the asymptotic value in the IR of $\sqrt{3M^3}<-\beta_\infty<2\sqrt{3M^3}$. From Eq.~\eqref{eq:beta_function} it is then clear that for solutions which satisfy this condition the IR value of the $\beta$-function is simply given by
\begin{equation}\label{eq:betainf}
 -\beta_\infty=\lim_{\phi\rightarrow\infty}\frac{3}{2}M^3\frac{V'(\phi)}{V(\phi)}=\frac{1}{2}\sqrt{3M^3\alpha}\left(1+\tanh\left(\frac{\gamma}{\nu}\right)\right).
\end{equation}
For $\gamma/\nu\gg1$ this leads once again to the requirement that $1\lesssim\alpha\lesssim4$.

\begin{figure}[h]
\begin{center}
\includegraphics[height=7cm]{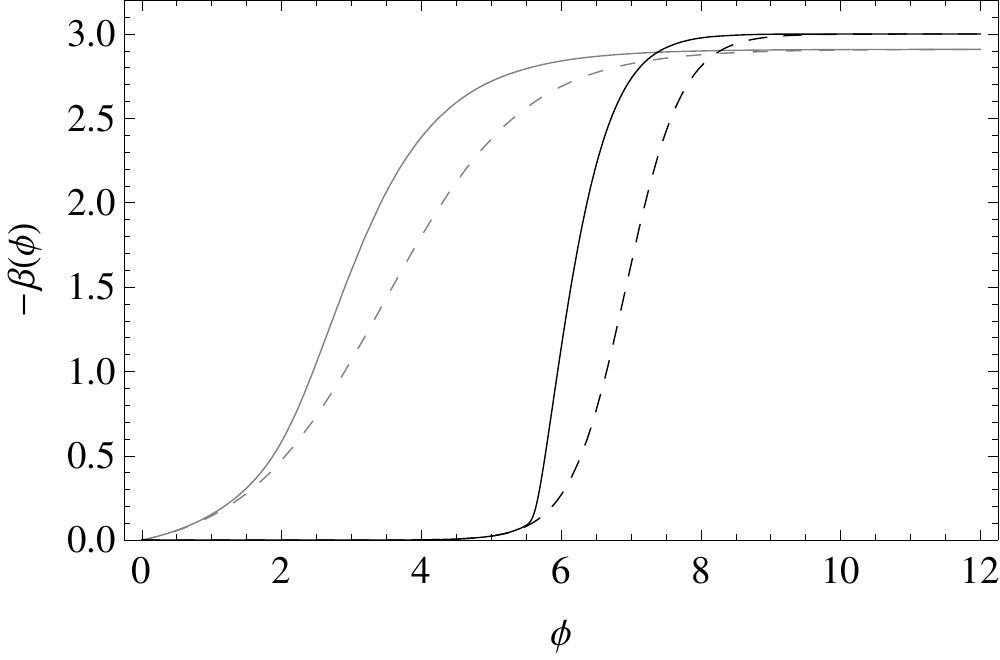}
\end{center}
\caption{The $\beta$-function as a function of $\phi$. The dashed lines show $\frac{3M^3}{2}\frac{V'}{V}$ for comparison. The black lines correspond to values of $\nu=0.8$ and $\gamma=4\sqrt{\alpha}$, while the grey lines are for $\nu=2$ and $\gamma=2\sqrt{\alpha}$. We have set $\alpha=3$ and $M=k=1$.}
\label{fig:numerical_beta}
\end{figure}

It is now straightforward to numerically solve Eq.~\eqref{eq:beta_function} by shooting from the above value in the IR. The $\beta$-function is shown for various parameter choices in Figure~\ref{fig:numerical_beta}. Notice in particular that for the solid black curve, corresponding to $\nu=0.8$, the $\beta$-function exhibits the behaviour described above. It is walking in the UV, corresponding to a nearly marginal deformation, followed by the turning on of a condensate and a sharp rise to confinement in the IR. This is in contrast to the $\beta$-function for our superpotential model in Figure~\ref{fig:beta} where there is no walking region and the condensate turns on at $\phi=0$.

Note also that when $\gamma=0$ the potential simplifies, but in this case values of $\alpha$ and $\nu$ which lead to a near-marginal operator result in a $\beta$-function with no clear transition region and a slow rise to confinement. It will become clear that this case, similarly to the potential in Eq.~\eqref{eq:general_potential1}, will not lead to a light dilaton.

\begin{figure}[h]
\begin{minipage}[b]{0.5\linewidth}
\begin{center}
\includegraphics[height=5.5cm]{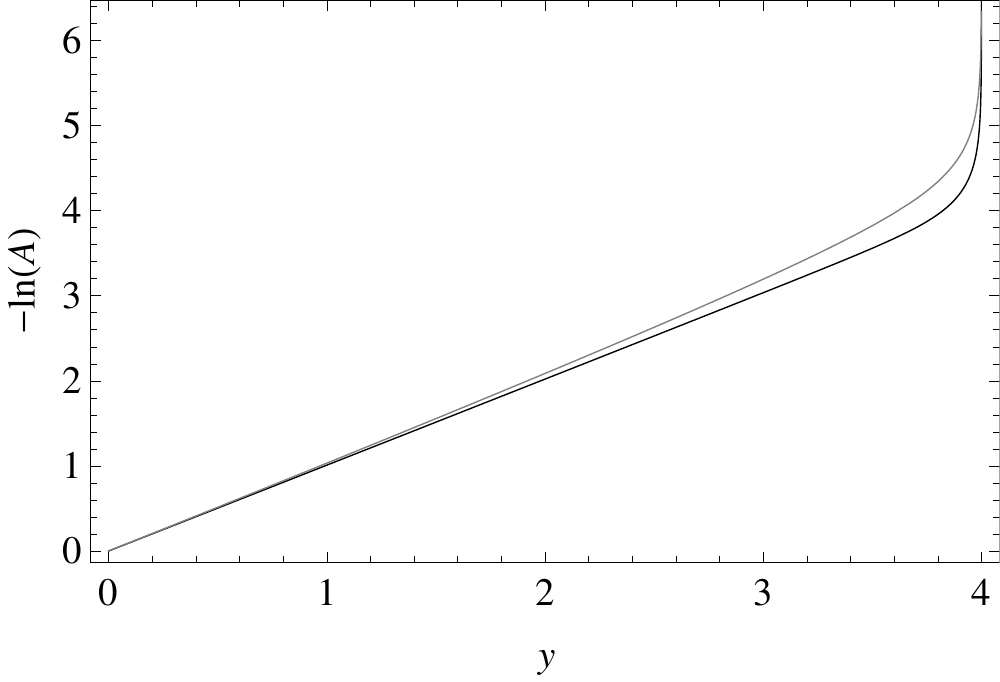}
\end{center}
\end{minipage}
\begin{minipage}[b]{0.5\linewidth}
\begin{center}
\includegraphics[height=5.5cm]{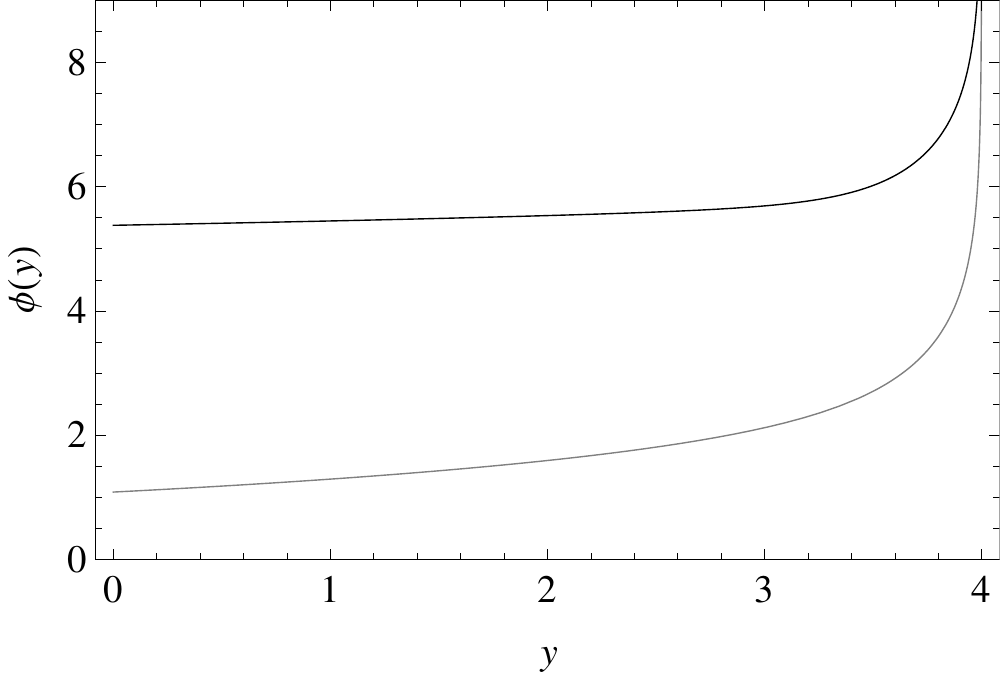}
\end{center}
\end{minipage}
\caption{The left and right panels show the solutions for the warp factor and bulk scalar profile respectively. The black lines correspond to values of $\nu=0.8$ and $\gamma=4\sqrt{\alpha}$, while the grey lines are for $\nu=2$ and $\gamma=2\sqrt{\alpha}$. We have set $\alpha=3$, $y_c=4$ and $M=k=1$.}
\label{fig:numerical_bkg}
\end{figure}

Now that we have a solution for the $\beta$-function, the background metric and scalar profile can easily be obtained by numerically integrating Eqs.~\eqref{eq:beta_def} and \eqref{eq:bulk_bkg_eom}. The solutions are shown in Figure~\ref{fig:numerical_bkg}. Notice that for the case of small $\nu$ the scalar profile is approximately constant in the UV, consistent with the fact that we have a nearly marginal operator with the dimension explicitly given by
\begin{equation}
[{\cal O}] \simeq 4 - \frac{2\alpha}{\nu} e^{-\frac{2\gamma}{\nu}}\,.
\end{equation}

%%%%%%%%%%%%%%%%%%%%%%%%%%%%%%%%%%%%%%%%%%%%%%%%%%%%%%%%%%%%%%%%%%%%%%%%%%%%%%%%%%%%%%%%%%%%%%%%%%%
\subsection{Dilaton mass}

We are now in a position to investigate the behaviour of the dilaton mass for our potential in Eq.~\eqref{eq:general_potential}. It was demonstrated in Ref.~\cite{Megias:2014iwa} that there are two conditions which should be satisfied in order for the dilaton to be light. Firstly, the $\beta$-function must remain small into the IR region where the condensate is turning on. This corresponds to the original statement in~\cite{CPR} that a light dilaton occurs for near-marginal operators. The second condition requires that the rise of the $\beta$-function to confinement is sufficiently fast. From Figure~\ref{fig:numerical_beta} it is clear that using our form for the bulk potential we are able to satisfy both of these conditions and thus expect to find a light dilaton in the spectrum. 

\begin{figure}[h]
\begin{minipage}[b]{0.5\linewidth}
\begin{center}
\includegraphics[height=5.3cm]{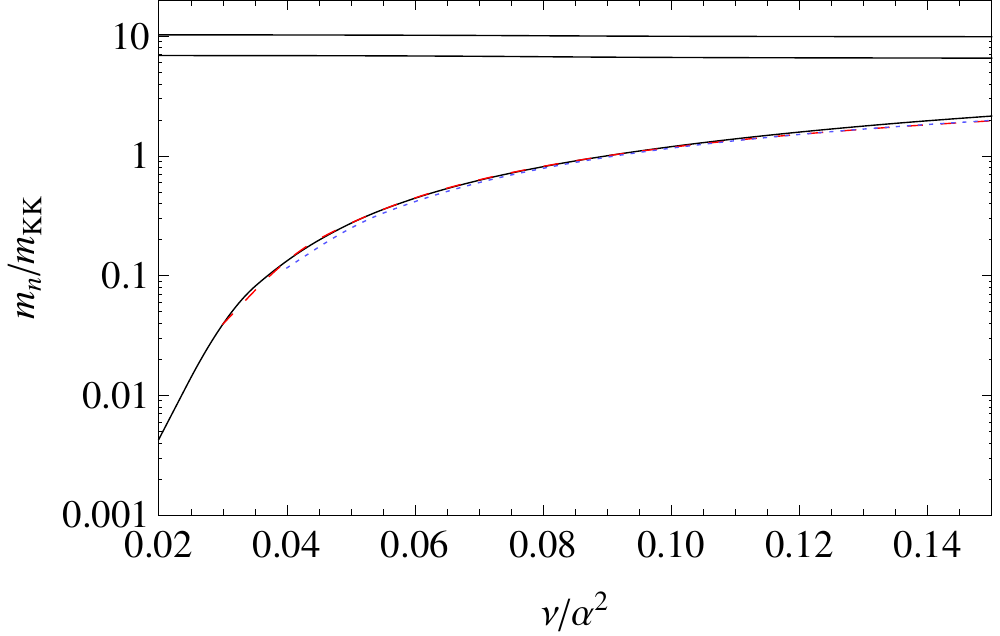}
\end{center}
\end{minipage}
\begin{minipage}[b]{0.5\linewidth}
\begin{center}
\includegraphics[height=5.45cm]{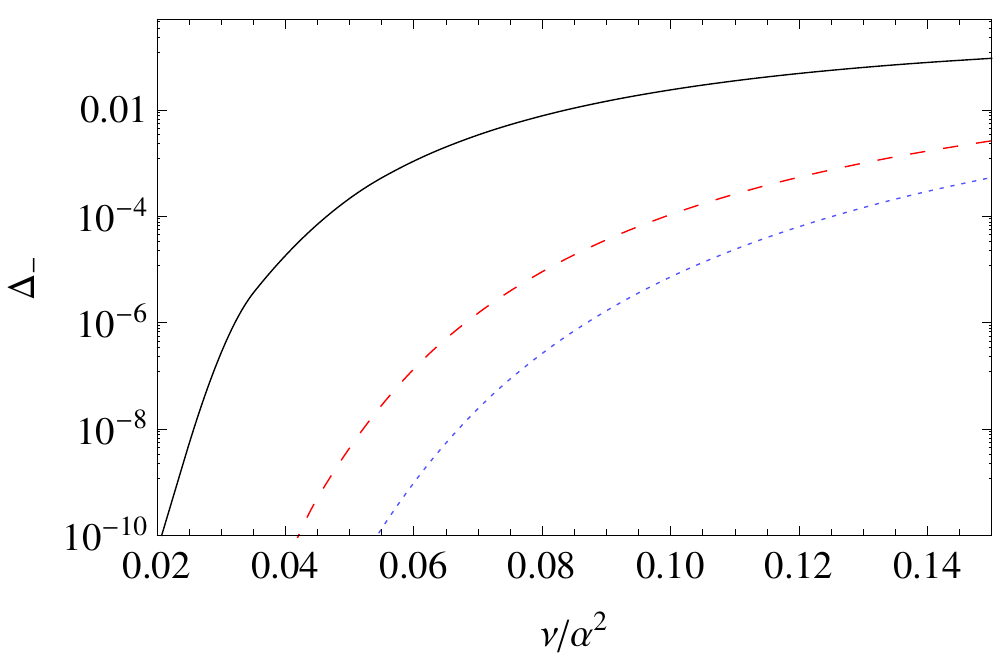}
\end{center}
\end{minipage}
\caption{Masses of the three lightest modes (left panel) and $\Delta_-$ (right panel) as a function of $\nu/\alpha^2$ for fixed values of $\alpha$ and $\gamma$. The solid black and dashed red lines are for $\alpha=3.8$ with $\gamma=2\sqrt{\alpha}$ and $4\sqrt{\alpha}$ respectively. The dotted blue lines are for $\alpha=2$ and $\gamma=2\sqrt{\alpha}$. We have set $ky_c=4$.}
\label{fig:dil_mass}
\end{figure}

Again solving Eq.~\eqref{eq:scalar_eom} in the $m_{UV}\rightarrow\infty$ limit we obtain the masses of the dilaton and two lightest KK modes, which are shown as a function of $\nu/\alpha^2$ in the left panel of Figure~\ref{fig:dil_mass}. Notice that the behaviour of the dilaton mass is almost independent of the choice of values for $\alpha$ and $\gamma$. This suggests that the ratio $\nu/\alpha^2$ determines both the value of the $\beta$-function where the condensate is turning on as well as the gradient in the transition region. The criterion for obtaining a light dilaton is then simply $\nu/\alpha^2\lesssim0.1$. We can see from Eq.~\eqref{eq:V'V} that the gradient of $V'/V$ in the transition region is determined by the ratio $\alpha/(2\nu)$ and so cases which satisfy our condition for a light dilaton correspond as expected to a fast transition between the walking UV and confining IR. 

In the right panel of Figure~\ref{fig:dil_mass} we have also plotted $\Delta_-$ as a function of $\nu/\alpha^2$. This clearly demonstrates that simply having a near-marginal operator is not on its own a sufficient condition to obtain a light dilaton. However it can be seen that, for our potential, cases where the transition to confinement is sufficiently fast will necessarily correspond to cases where we have a near-marginal operator. We also point out that although $\Delta_-$ is clearly dependent on $\gamma$, by looking again at Eq.~\eqref{eq:V'V} we see in the limit $\gamma/\nu\gg1$ that $\gamma$ simply corresponds to a shift in $\phi$ and so is not expected to have an effect on the dilaton mass. This highlights the fact that in order to obtain a light dilaton it is the $\beta$-function which must remain small at the condensate scale, rather than just $\Delta_-\ll1$. 

Finally, we comment briefly on the naturalness of our model. Since we are only able to solve the equations of motion numerically for the bulk potential in Eq.~\eqref{eq:general_potential}, it is difficult to directly construct the dilaton effective potential as in Section~\ref{sec:effective_potential} and verify the existence of a non-zero quartic term for the dilaton. However, a closer look at our solution suggests that this construction is indeed natural. Firstly, it is clear from Figure~\ref{fig:numerical_beta} that we have a $\Delta_-$-type solution in the parameter regions of interest. In fact this is guaranteed once we require that the $\beta$-function is walking in the UV. Secondly, near-marginal operators do not admit an alternative CFT description and therefore tuned $\Delta_-$-type solutions (with $\xi=0$ and zero quartic potential) describe deformed CFTs with no condensate. These tuned solutions are not expected to give rise to a light mode except when (i) $\Delta_-\rightarrow2$, (ii) if a tuning is directly imposed between the bulk and boundary masses for a finite UV cut-off, or (iii) in cases such as that shown in Ref.~\cite{Megias:2014iwa} where $V'/V$ deviates significantly from the $\beta$-function in the transition region as a result of the potential being carefully chosen to avoid the formation of a condensate. Clearly none of these situations are relevant for our scenario. We therefore conclude that our model does indeed correspond to the case described in~\cite{CPR} whereby a large quartic coupling in the UV is driven towards zero at some hierarchically small scale by the running of a near-marginal operator, leading to spontaneous breaking of conformal symmetry and a light dilaton.

%%%%%%%%%%%%%%%%%%%%%%%%%%%%%%%%%%%%%%%%%%%%%%%%%%%%%%%%%%%%%%%%%%%%%%%%%%%%%%%%%%%%%%%%%%%%%%%%%%%
\section{Conclusion}\label{sec:conclusion}

We have presented two solutions of the Einstein equations that exhibit the properties of a dilaton in a soft-wall background. The first solution, based on the superpotential method, describes a CFT with conformal symmetry broken by a finite-dimensional operator with dimension $1 < [{\cal O}] < 4$. This represents a more realistic scenario compared to hard-wall models that are dual to theories with an infinite-dimensional operator. When the symmetry is spontaneously broken, a light dilaton appears in the spectrum provided there is a hierarchy between the UV and IR scales. In the case of explicit breaking a light dilaton is obtained when the operator dimension is near two. However in both cases the quartic coupling of the effective dilaton potential is tuned and requires an additional symmetry, such as supersymmetry, to be naturally realised. This is not surprising since the solution is based on an analytic superpotential which does not contain a nonperturbative term.

The second solution generalises the bulk scalar potential obtained from the superpotential method to allow for non-analytic terms in the $\beta$-function. The bulk scalar potential depends on three parameters that controls the form of the $\beta$-function from the UV to the IR scale. Interestingly it is possible to obtain a $\beta$-function that remains approximately constant from the UV scale until the condensation scale, at which point the $\beta$-function rises sharply to an approximately constant order-one value at the IR scale. Our parameterisation relates the slope of this fast rise precisely to the near-marginality of the operator dimension. In this case we explicitly find a naturally light dilaton, which agrees with the results obtained in Ref.~\cite{Megias:2014iwa}. Our solution provides a simple example of how to obtain a naturally light dilaton from nearly-marginal CFT deformations in a soft-wall background. It would be interesting to study the underlying dual theory that is responsible for the particular bulk scalar potential. This could then be useful for applying these ideas to electroweak symmetry and the Higgs sector in the Standard Model.

%%%%%%%%%%%%%%%%%%%%%%%%%%%%%%%%%%%%%%%%%%%%%%%%%%%%%%%%%%%%%%%%%%%%%%%%%%%%%%%%%%%%%%%%%%%%%%%%%%%
\section*{Acknowledgements}

We thank Alex Pomarol for helpful discussions. This work was supported by the Australian Research Council. TG was also supported by the Department of Energy grant DE-SC0011842 at the University of Minnesota. TG thanks the hospitality of the Aspen Center for Physics and the partial support from National Science Foundation Grant No\@. PHYS-1066293 where part of this work was done. PC is grateful to the School of Physics and Astronomy at the University of Minnesota for its hospitality during the completion of this work.

%%%%%%%%%%%%%%%%%%%%%%%%%%%%%%%%%%%%%%%%%%%%%%%%%%%%%%%%%%%%%%%%%%%%%%%%%%%%%%%%%%%%%%%%%%%%%%%%%%%
\bibliographystyle{biblio_style}
\bibliography{soft_wall_dilaton}

\end{document}